# Gender Imbalance in Physics Education and Employment in Germany: Trends and Challenges

Ruzin Aganoglu[1, 7, a)], Andrea Bossmann[1, 2, 3, c)], Ulrike Böhm[1, 8, g)],
Anja Metzelthin[5, e)], Agnes Sandner[1, b)], Iris Traulsen[1, 6, f)],
and Angelica Zacarias[1, 4, d)]

[1]*Working Group on Equal Opportunities of the German Physical Society, Hauptstr. 5, 53604 Bad Honnef, Germany*
[2]*Department of Physics, Freie Universität Berlin, Arnimallee 14, 14195 Berlin, Germany*
[3]*Department of Humanities and Educational Sciences, Center for Interdisciplinary Women's and Gender Studies, Technische Universität Berlin, Fraunhoferstr. 33-36, 10587 Berlin, Germany.*
[4]*Max Planck Institute of Microstructure Physics, Weinberg 2, 06120 Halle, Germany*
[5]*German Physical Society, Hauptstr. 5, 53604 Bad Honnef, Germany*
[6]*Leibniz-Institut für Astrophysik Potsdam (AIP), An der Sternwarte 16, 14482 Potsdam, Germany*
[7]*Lab-on-Fiber GmbH, Pommernstraße 32, 96242 Sonnefeld, Germany*
[8]*Carl Zeiss AG, Carl-Zeiss-Strasse 22, 73447 Oberkochen, Germany*

*Author Emails*
[a)] Corresponding author: ruzin@physik.fu-berlin.de
[b)] sandner@akc.dpg-physik.de
[c)] a.bossmann@fu-berlin.de
[d)] zacarias@mpi-halle.mpg.de
[e)] metzelthin@dpg-physik.de
[f)] itraulsen@aip.de
[g)] boehm@akc.dpg-physik.de

**Abstract.** Gender imbalance among German physicists persists, with fewer women in advanced degrees and research leadership roles. Although female enrollment in Physics programs increased slightly until 2022, potentially influenced by COVID-19 the long-term trend remains uncertain. Despite the rise in female Ph.D. students and foreign representation, female professors in Physics and Astronomy stagnated below 14 %, indicating significant underrepresentation. Anticipated revisions to the WissZeitVG law may impact female mainstreaming efforts, potentially leading to greater precarization of research staff. Women make up only around 20 % of employed physicists, with low visibility in the community, as seen in the representation of female physicists in prestigious awards. Addressing this imbalance requires structural interventions beyond mere encouragement and empowerment.

## ENROLLMENT IN PHYSICS STUDIES

Data indicate that the percentage of female students enrolled in Bachelor's Physics programs remained constant at 35 % between 2020 and 2022, whereas their share was smaller at approximately 23 % during Master's studies. This suggests a gender imbalance in Physics, with fewer women pursuing advanced degrees. However, it is worth noting that the percentage of female students in both programs has increased slightly over the years. Female enrollment in Bachelor's programs have tended to decrease now compared to 2019, when more than 39% of the B.Sc. Physics students were female [1]. This trend could be linked to the 6.5% decrease in the science, technology, engineering, and mathematics (STEM) field enrollment in Germany due to the recent COVID-19 crisis [2]. However, current data are insufficient to predict whether it is a temporary or a longer-term effect [3].

The number of female Ph.D. students in Physics has gradually increased, from 2,079 (23 %) in 2019 to 2,310 (24 %) in 2021. Among the 2,310 females pursuing their Ph.D. in 2021, the share of foreign female students increased by 3 % compared to 2019, making up 47 %. On the contrary, the proportion of male foreign Physics Ph.D. students has not increased as much as that of female students, standing at 29 % in 2021 [4]. This trend implies that there may be a higher demand for female Physics Ph.D. students from foreign countries to study, whereas the native population has fewer female Ph.D. candidates than males. Overall, involving more female students from abroad may improve diversity in Physics, particularly in Ph.D. programs. A more in-depth analysis of the reasons behind this shift and its potential impact on the field requires additional data.

## RESEARCH LEADERSHIP

In Physics and Astronomy at German universities, the number of women among professors was 164 (12.9 %) in 2019 and 176 (13.7 %) in both 2020 as well as 2021 [5]. This indicates a significant underrepresentation of women in research leadership positions. The data indicate an increasing percentage of female professors over the last years, but the pace of change has been slow. For example, the percentage of female professors in Physics and Astronomy has only increased by 2.8 % since 2015. The low representation of women in research leadership positions in Physics and Astronomy is a well-known issue in Germany similar to that in many other countries. It arises from various factors, including systemic biases and gender stereotypes, that can create barriers to entry and promotion of women in science. The upcoming revision of the German law that regulates the employment of scientific staff in universities and research institutions, known as the "Wissenschaftszeitvertragsgesetz" (WissZeitVG) [6], is expected to be controversial in terms of its impact on female mainstreaming efforts. It is anticipated that the revision will allow for more fixed-term contracts, which could contribute to the precarization of research staff, particularly for early-career researchers and women. An increase in fixed-term contracts can create job insecurity and limit career opportunities, disproportionately affecting women who may face additional challenges in terms of work-life balance, parental leave, and career breaks. This exacerbates the "leaky pipeline", wherein women are even less likely to advance to higher positions and are more likely to eventually leave the academic workforce. The Working Group on Equal Opportunities, in parallel with the efforts of the German Physical Society [7], lobbies and strives to achieve more appropriate work conditions for women and all researchers in Physics.

## OCCUPATION AND VISIBILITY

In 2012, the number of female physicists with employment liable to social insurance was 2,409. By 2021, this figure had grown to 3,349, which is a positive development. Nonetheless, it is concerning that only 20 % of the employed physicists are female, revealing a gender gap in physics employment [8]. The number of socially insured employees includes only those whose occupational field is explicitly declared as 'physicists' in the official labor statistics. There are also physicists whose occupational fields are declared as project manager, product engineer, etc. Nevertheless, the ratio of male to female can be extrapolated to all physicists, as can the general increase.

The representation of women in the Physics community has also been inadequate. Among the 14 esteemed German Physical Society (DPG) Physics awards, female physicists account for only 5 % of award recipients in the German Physics community on an average [9]. Thus, further action may be necessary to address this imbalance to inspire and support women in pursuing careers in Physics and to implement positive measures toward higher diversity at work and in academia.

## NETWORKING FOR GENDER EQUALITY IN PHYSICS

The Working Group on Equal Opportunities (AKC) of the German Physical Society is dedicated to addressing gender disparity in Physics and advancing equal opportunities for women and men. Its recent initiatives include workshops covering various topics such as entrepreneurship, time management, and achieving a balance between family life and career. Additionally, since its inception in 1997, the AKC has organized the annual Women in Physics conference held in different cities across Germany, bringing together over 200 women each year to discuss physics and societal issues [10].

The AKC has also pioneered other significant endeavors within the German Physical Society such as the establishment of the Hertha Sponer Prize in 2002, which honors outstanding young female physicists. Furthermore, the group conducts surveys to assess the status of women in Physics in Germany.

In pursuit of promoting physics education, the AKC initiated the school-level outreach program "Faszination Wissenschaft! MINT-Role-Models" (Fascinating Science! STEM Role Models) in collaboration with the German

Chemical Society (GDCh), Halles Schülerlabor für Physik, Martin Luther University Halle-Wittenberg (MLU), and Max Planck Institute for Microstructure Physics. This program offers activities for students from grades 6 through 13 (age 10 to 18 years), including seminars and webinar series featuring experienced scientists. These webinars are accessible online [11] and aim to showcase successful female STEM professionals, thereby bridging the gender gap in STEM fields.

Since January 2018, the AKC has regularly highlighted women in Physics through its "Physikerin der Woche (Woman in Physics of the Week)" initiative [12], publishing short portraits of female physicists who work in Germany or are German and work abroad. It aims at enhancing the visibility of women in Physics.

## CONCLUSION

In summary, the situation of female physicists in Germany reveals persistent gender imbalances, particularly in research leadership positions and visibility. Although progress has been recorded in increasing the number of female Physics Ph.D. students and employed physicists, challenges remain. Addressing systemic biases and structural barriers is crucial to cope with these challenges. Efforts should also focus on promoting equitable opportunities, enhancing visibility, and fostering an inclusive environment to inspire and support Physics careers of women. Collaboration among stakeholders is essential to enact comprehensive strategies for a more diverse and equitable future in the field.